\documentstyle[twoside,fleqn,espcrc2,epsf]{article}

\title{Lattice Gross-Neveu model 
with domain-wall fermions\thanks{presented by K.-i. Nagai}}

\author{Taku Izubuchi  
\address{Institute of Physics,  University of Tsukuba,
         Tsukuba, Ibaraki 305-8571, Japan}
and Kei-ichi Nagai 
\address{Center for Computational Physics,  University of Tsukuba,
         Tsukuba, Ibaraki 305-8577, Japan}}

\begin{document}
\pagestyle{empty}

\begin{abstract}
We investigate 
the two-dimensional lattice Gross--Neveu model,
using the domain-wall fermion formulation,
as a toy model of lattice QCD.
We study features of the phase diagram related to 
the mechanism of chiral symmetry restoration,
and find that the parity-broken phase (Aoki phase) 
exists for finite extent in the extra dimension ($N_s$). 
We also find that $O(a)$ scaling violation terms 
vanishes in the limit of $N_s\to\infty$.
\end{abstract}

\maketitle

\section{Introduction}

Defining chiral fermions on the lattice has been  
one of long-standing problems 
in lattice field theories.
Several years ago, 
domain-wall fermion \cite{domain1,domain2} has been proposed
as a new formulation of lattice chiral fermion.
This formulation considers Wilson fermion in $D+1$ dimensions 
with the free boundary condition in the extra dimension of a size $N_s$,
or equivalently, $N_s$-flavored Wilson fermions with flavor mixing.
In the limit $N_s\rightarrow\infty$, the spectrum of 
free domain-wall fermion (DWF) contains massless modes
at the edges in the extra dimension. 
The massless modes are stable under perturbation from weak gauge fields.

While there have been numerical simulations
of lattice QCD with DWF (DWQCD),
some non-perturbative issues,
in particular existence of the parity-broken phase (Aoki phase)\cite{saoki}
and necessity of fine tuning of couplings to restore chiral symmetry
for finite $N_s$,
have not been clarified.
In order to answer these questions,
we investigate the two-dimensional lattice Gross--Neveu model using 
the DWF formalism (DWGN) \cite{DWGN} as a test of DWQCD.

\section{Action and Effective potential}
We propose the following action for DWGN:
\begin{eqnarray}
&&S 
= S_{free} + a^2 \sum_{n}\bar{q}(n) 
\left\{ \sigma(n) + i \gamma_5 \Pi(n) \right\} q(n)
\nonumber \\
&&+ a^2 \sum_{n} \left[ \frac{N}{2 g_{\sigma}^2} 
\left\{\sigma(n) - m_f \right\}^2 + \frac{N}{2 g_{\pi}^2} \Pi(n)^2 
\right].
\label{eq:GNterm}
\end{eqnarray}
The auxiliary fields are related to fermion condensates 
as 
$ \sigma(n) =  m_f - \frac{\displaystyle{g_{\sigma}^2}}{\displaystyle{N}}
 \bar{q}(n)q(n)$ and 
$ \Pi(n) =  - \frac{\displaystyle{g_{\pi}^2}}{\displaystyle{N}} 
\bar{q}(n) i \gamma_5 q(n) $.
The important point in this action
is that the interaction terms are constructed 
from the edge state: 
$q(n) = P_R \psi(n,s=1) + P_L \psi(n,s=N_s)$.

We attempt to solve DWGN analytically in the large $N$ limit.
The effective potential can be calculated 
by the technique of the propagator matrix\cite{truncate1,truncate2}, 
which yields
\begin{eqnarray}
&& V_{eff} = \frac{1}{2 g_{\sigma}^2} \left( \sigma - m_f \right)^2
 +  \frac{1}{2 g_{\pi}^2} \Pi^2 - I  ~,
\label{eq:pot} \\
&& I = \int_p
\ln \left[ F a^2 (\sigma^2 + \Pi^2) + G a \sigma + H
\right] ~,
\label{eq:effpot}
\end{eqnarray}
where $\int_p = \int_{-\pi/a}^{\pi/a} 
\frac{d^2 p}{(2 \pi)^2}$.
The explicit form of the functions $F, G$ and $H$
are described in Ref.\cite{DWGN}.
Let us note that the term $G a \sigma$ 
explicitly breaks chiral symmetry.

\section{Existence of Aoki phase}

\begin{figure}[t]
\centerline{\epsfxsize=7.5cm \epsfbox{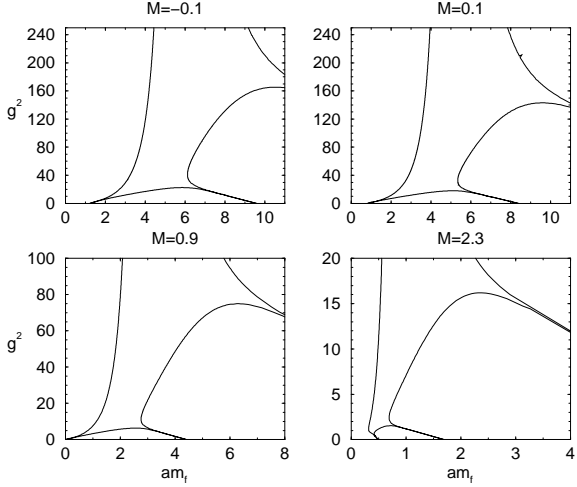}}
\vspace{-1cm}
\caption{
Phase diagram at $N_s = 2$ on ($a m_f$ , $g^2$) plane
for several values of $M$.}
\label{fig:Ns2}
\end{figure}

\begin{figure}[t]
\centerline{\epsfxsize=6.5cm \epsfbox{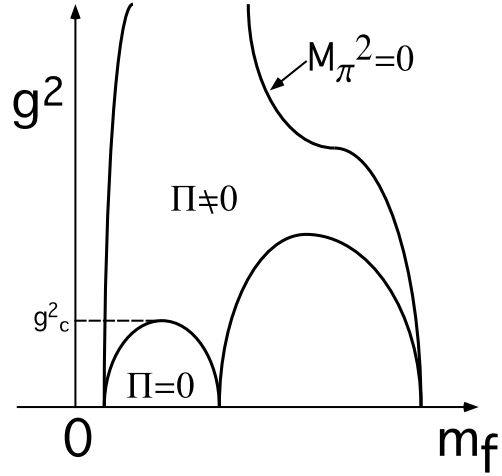}}
\vspace{-1cm}
\caption{
Schematic phase diagram 
on $(m_f,g^2)$ plane for $N_s=$ even.
}
\label{fig:illust1}
\end{figure}

The phase diagram of the Wilson fermion action
has a region of spontaneously broken parity-flavor 
symmetry (Aoki phase), which plays an important role for 
controlling restoration of chiral symmetry in the continuum 
limit\cite{saoki,noaki}.  Since DWF is an extension of the Wilson 
fermion formalism, we wish to examine if Aoki phase exists 
for DWGN.
In this section, we set the coupling constants as $g_\sigma^2 = g_\pi^2 = g^2$.

Let us first consider the case of $N_s=\infty$.
Since $F$ and $H$ dominates over the term $G$ in this limit,
the effective potential becomes 
\begin{equation}
I(\sigma,\Pi) = \int_p \ln \left[
F a^2 (\sigma^2 + \Pi^2 ) + H \right]~.
\end{equation}
The $O(a)$ term, which breaks chiral symmetry, is absent, 
and hence the model has exact chiral symmetry.  
Thus pion becomes massless without fine tuning 
even for finite lattice spacings 
and for arbitrary strong coupling.

Next we consider the case of finite $N_s$.
It is expected that Aoki phase exists in this
case since the $N_s=1$ DWGN is equivalent to 
GN model with Wilson fermion\cite{noaki}.

The solution of the gap equation marking the phase boundary 
is illustrated for several values of $M$ with 
a fixed size $N_s=2$ in Fig.\ref{fig:Ns2}.  
As is summarized in a schematic diagram in Fig.\ref{fig:illust1},  
we find that
(i) Aoki phase exists inside the boundary
and this boundary forms cusps toward $g^2\to 0$, 
(ii) the Aoki phase always appears in the $m_f > 0$ region
for even $N_s$ (in the conventional choice of sign in domain-wall
literatures, this corresponds to $m_f<0$), 
(iii) pion becomes massless on the boundary.
The last point means that pion mass does not vanish at $m_f=0$ for finite
lattice spacing.  Hence a fine tuning is needed to obtain massless pion.

\begin{figure}[t]
\centerline{\epsfxsize=7cm \epsfbox{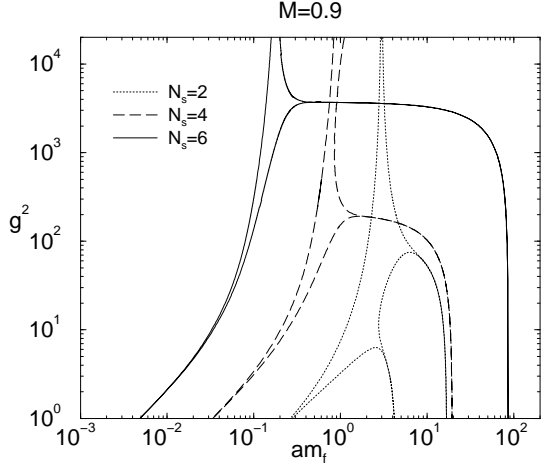}}
\vspace{-1cm}
\caption{
$N_s$ dependence of the phase boundary
}
\label{fig:MNs}
\end{figure}

In Fig.\ref{fig:MNs} we show the $N_s$ dependence of the phase boundary
for $N_s=2, 4$ and $6$ at $M=0.9$.
We find that 
(i) the Aoki phase shrinks exponentially with $N_s$, 
(ii) the first ``finger'' on the left approaches $m_f=0$, 
while other ``fingers'' move toward $m_f=\infty$, 
(iii) the critical value $g_c$ (see Fig.\ref{fig:illust1}), 
below which the cusp structure (``fingers'') appears,
increases exponentially with $N_s$,
in contrast to the case of Wilson fermion\cite{noaki}.
These features mean that the area of the normal phase becomes wide
with increasing $N_s$.

As seen from above, DWF for finite $N_s$ represents an improved 
Wilson fermion.

\section{$O(a)$ scaling violation ($a \rightarrow 0$)}

\begin{figure}[t]
\centerline{\epsfxsize=7.5cm \epsfbox{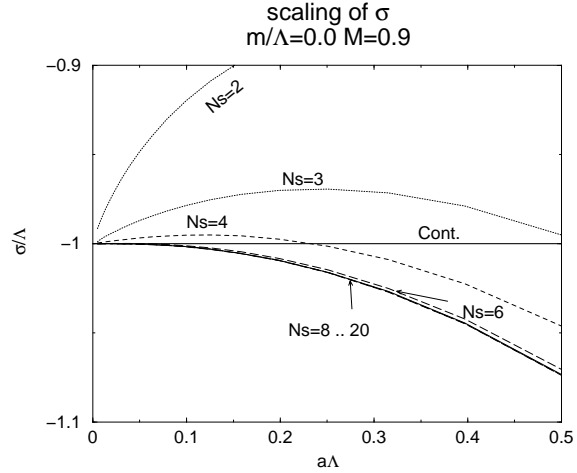}}
\vspace{-1cm}
\caption{Scaling of $\sigma/\Lambda$ as a function of $a \Lambda$}
\label{fig:Obs_a-Mfix}
\end{figure}

We study the mechanism of chiral symmetry restoration
in the continuum limit.
In particular, we wish to understand scaling violation
for finite and infinite $N_s$.

The effective potential in the continuum limit
is obtained after some calculation:
\begin{equation}
V_{eff} = - m  \sigma_R
+ \frac{1}{4 \pi} \left( \sigma_R^2 + \Pi_R^2 \right)
\ln \frac{\sigma_R^2 + \Pi_R^2}{e \Lambda^2}~.
\label{eq:effpot1}
\end{equation}
Here 
$\sigma_R = f_M 
\{\sigma - \frac{\displaystyle{(1-M)^{N_s}}}{\displaystyle{a}}\}$ 
and $\Pi_R = f_M \Pi$
with 
$f_M = \frac{\displaystyle{M(2-M)}}{\displaystyle{1-(1-M)^{2 N_s}}}$,
which is the normalization factor of the edge state ``$q(n)$''
for finite $N_s$.
In order to obtain (\ref{eq:effpot1}), 
which agrees with the continuum result,
we need to impose the following scaling relations:
\begin{eqnarray}
&&\frac{1}{2 g_\sigma^2} - \hat{C_0} + C_2 
= \frac{f_M^2}{4 \pi} \ln \frac{1}{a^2 \Lambda^2} ~, 
\label{eq:gsgtuning}\\
&&\frac{1}{2 g_\pi^2} - \hat{C_0}  
= \frac{f_M^2}{4 \pi} \ln \frac{1}{a^2 \Lambda^2} ~, 
\label{eq:gpituning}\\
&&\frac{1}{f_M} 
\left( \frac{m_f}{g_\sigma^2} - \frac{(1-M)^{N_s}}{g_\sigma^2 a} 
+ \frac{C_1}{a} \right) = m ~.
\label{eq:tuning}
\end{eqnarray}
These relations are the same with those found for the Wilson fermion case 
\cite{noaki}.
Therefore a fine tuning is needed to restore chiral symmetry.

Figure~\ref{fig:Obs_a-Mfix} shows 
$\sigma/\Lambda$ as a function of $a \Lambda$
for $M=0.9$,
using the Wilson-like scaling relations
in (\ref{eq:gsgtuning}-\ref{eq:tuning}).
We find that $O(a)$ scaling violation is large at $N_s=2,3$.
However, the magnitude of $O(a)$ scaling violation diminishes exponentially as $N_s$ 
increases.  In fact the scaling curve almost exactly follows the $O(a^2)$ 
behavior for $N_s\geq 8$.

\section{Conclusions}

We have investigated the two-dimensional DWGN model
in detail as a test of DWQCD.
When the size of the extra dimension $N_s=\infty$, 
the model has chiral symmetry for finite lattice spacing, 
and Aoki phase does not exist.  On the other hand, 
in the case of finite $N_s$, Aoki phase does exist
and a fine tuning is needed to restore chiral symmetry in 
the continuum limit.
However, the $O(a)$ scaling violation that gives rise to 
this behavior vanishes exponentially fast as $N_s$ is increased 
so that it is negligible in practice for $N_s=O(10)$.


While the GN model does not have gauge fields and quantum 
fluctuations are absent in the large $N$ limit,  
it is expected that the results obtained in this work
provide instructive and systematic information
for DWQCD simulations.

The authors are JSPS Research Fellows.

%

%
%

\end{document}